\documentclass[doublecol]{epl2} 

\begin{document}

\def\jpb{J. Phys. B: At. Mol. Opt. Phys.~}
\def\pra{Phys. Rev. A~}
\def\prb{Phys. Rev. B~}
\def\prl{Phys. Rev. Lett.~}
\def\jmo{J. Mod. Opt.~}
\def\jetp{Sov. Phys. JETP~}
\def\etal{{\em et al.}}

\def\reff#1{(\ref{#1})}

\def\diff{\mathrm{d}}
\def\imagi{\mathrm{i}}

\def\beq{\begin{equation}}
\def\eeq{\end{equation}}

\def\ket#1{\vert #1\rangle}
\def\bra#1{\langle#1\vert}
\def\braket#1#2{\langle #1 \vert #2 \rangle}

\def\expect#1{\langle #1 \rangle}

\def\vekt#1{\vec{#1}}
\def\vect#1{\vekt{#1}}
\def\vektr{\vekt{r}}

\def\makered#1{{\color{red} #1}}

\def\Im{\,\mathrm{Im}\,}
\def\tr{\mathrm{tr}}
\def\dspin{\hat{\bar{\psi}}}
\def\spin{\hat{\psi}}

\def\Ad{\hat{\dot{A}}}
\def\A{\hat{A}}

\newcommand{\be}{\begin{equation}}
\newcommand{\ee}{\end{equation}}
\newcommand{\bea}{\begin{eqnarray}}
\newcommand{\eea}{\end{eqnarray}}
\newcommand{\br}{\mathbf{r}}

\title{Global fixed point proof of
time-dependent density-functional theory}

\author{M Ruggenthaler \and R van Leeuwen}
\institute{Department of Physics, Nanoscience Center, University of Jyv\"askyl\"a, 40014 Jyv\"askyl\"a, Finland}


\abstract{
We reformulate and generalize the {uniqueness and existence proofs} of time-dependent density-functional theory. The central idea is to restate the fundamental one-to-one correspondence between densities and potentials as a global fixed point question for potentials on a given time-interval.  We show that the unique fixed point, i.e. the unique potential generating a given density, is reached as the limiting point of an iterative procedure. {
The one-to-one correspondence between densities and potentials is a straightforward result provided that the
response function of the divergence of the internal forces is bounded. The existence, i.e. the v-representability of a density, can be proven as well provided that the operator norms of the response functions of the members of the iterative sequence of potentials have an upper bound. The densities under consideration have second time-derivatives that are required to satisfy a condition slightly weaker than being square-integrable.} This approach avoids the usual restrictions of Taylor-expandability in time of the {uniqueness} theorem by Runge and Gross [\prl \textbf{52}, 997 (1984)] and {of the existence theorem by} van Leeuwen [\prl {\bf 82}, 3863 (1999)]. Owing to its generality, the proof not only answers basic questions in density-functional theory but also has potential implications in other fields of physics.}
\maketitle

Time-dependent density-functional theory (TDDFT) provides an exact reformulation of the time-dependent quantum many-body problem
in terms of an effective non-interacting problem that is functionally dependent on the one-particle density.
Since the effective problem is numerically much more tractable than the original one the method has found widespread use in
different fields of physics and chemistry \cite{TDDFT, vanLeeuwenIJMP, Marques}. The cornerstone of TDDFT is the existence of a one-to-one mapping between
densities and external potentials for a given initial state. The standard proofs that establish this mapping by Runge and Gross \cite{RG} 
and van Leeuwen \cite{vanLeeuwen} restrict the set of allowed potentials to those that have a Taylor-expansion around the initial time.
In this work we extend the{se proofs} to allow for general finite potentials without restrictions on their time-dependence. Hence we
can, for example, include potentials of the temporal form $\sqrt{t}$ or $\exp{(-1/t)}$ around $t=0$.
 This was until now only
proven to be allowed for the linear response case \cite{vanLeeuwenIJMP}, dipole fields \cite{RuggenthalerPRA} or lattice systems \cite{TokatlyArXiv}. 
{ÊThe proof of existence \cite{vanLeeuwen} also requires Taylor-expandability of the density. 
This issue is discussed for the one-dimensional case in \cite{Maitraetal}. In the present formulation we do not need this
property. The second time-derivative of the density is only required to satisfy a condition which is slightly weaker than being square-integrable.}
The central idea to go beyond the Taylor-expandability restriction{s} is to reformulate the fundamental one-to-one correspondence between densities and potentials 
as a global fixed point question for potentials on a given time-interval. We prove that the fixed point, i.e. the potential 
that generates a given density, is unique and exists under certain assumptions. \\
Let us outline the general problem.
We consider a many-body Hamiltonian {in atomic units} of the form
\begin{eqnarray}
\label{Hamiltonian} \hat{H}([v],t) =  \hat{T} + 
\hat{V}([v],t) + \hat{W} 
\end{eqnarray}
where the operator $\hat{T} = -\sum_{i=1}^N \frac{1}{2} \nabla_i^{2}$ is the kinetic term, $\hat{W}=\frac{1}{2}\sum_{i \neq j}^N  w(|\br_i-\br_j|)$ is the two-particle interaction,
\begin{eqnarray}
\label{ConjugateVaribales}
\hat{V}([v],t) = \sum_{i=1}^N v(\br_i t) = \int d \br \,\hat{n}(\br) v(\br t)
\end{eqnarray}
is the external potential and 
\begin{equation}
\hat{n}(\br)=\sum_{i=1}^N \delta (\br-\br_i)
\end{equation} 
is the density operator.
{The interaction potential $w (|\br-\br'|)$ can be any locally square-integrable potential with a Coulomb 
singularity or less singular, though it is usually assumed to be purely Coulombic.}
For a given initial state $\ket{\Psi_0}$ at time $t_0$ the solution of the time-dependent Schr\"odinger equation (TDSE) 
\begin{equation}
i \partial_t \ket{\Psi ([v],t)} = \hat{H}([v],t) \ket{\Psi ([v],t)} 
\end{equation}
for a potential $v$ (including gauge fixing) provides a mapping from potentials to wave-functions
$v \mapsto \ket{\Psi([v],t)}$.  {The wave-functions are as usually assumed to be square integrable and in the self-adjoint domain of the Laplacian, i.e. $\nabla^2 \psi$ is also square integrable.}
{Due to the mapping of potentials onto wave functions} the expectation value of any operator $\hat{O}$ becomes a functional of the external potential, i.e.
\begin{equation}
 O([v],t) = \braket{\Psi([v],t)| \hat{O}}{\Psi([v],t)}.
\end{equation}
In particular the density $n([v], \br t)$ is a functional of the potential and the initial state.
The central statement on which TDDFT is based is that this mapping
 $v \mapsto n$ is one-to-one and thus invertible, i.e. for a given density there is only one external potential generating this density 
 by solution of the TDSE for a given initial state $\ket{\Psi_0}$. 
 This implies that the wave-function and all observables can be considered as functionals of the density instead of the potential. 
 The statement is independent of the initial state and the two-particle interaction. This opens up the 
 possibility that the density of an interacting system can be reproduced by a unique effective potential in a non-interacting system.
 This forms the basis of the Kohn-Sham scheme that makes TDDFT computationally practical.
 Before we outline the general proof we reinvestigate the following
equation, which is of fundamental importance to TDDFT \cite{TDDFT, vanLeeuwen}
\begin{eqnarray}
\label{InitialSturmLiouville}
 -\nabla  \cdot  \left[ n([v],\br t) \nabla v(\br t) \right]   =   q([v], \br t) - \partial^{2}_t n([v], \br t) .
\end{eqnarray}
 It
 can be obtained by applying the Heisenberg equation of motion for the density operator twice.
In this equation 
\be
q([v], \br t) = \braket{\Psi([v],t)| \partial_{l}(\partial_{k} \hat{T}_{k l}(\br) +  \hat{W}_l (\br))}{\Psi([v],t)}
\label{qfunction}
\ee
where summing over multiple indices is implied and
\bea
\hat{T}_{kl} (\br) &=& \frac{1}{2} \Big( \sum_{i=1}^N \, \overleftarrow{\partial}_{k,i} \delta (\br -\br_i) \overrightarrow{\partial}_{l,i} + (k \leftrightarrow l) \Big) \nonumber \\
&-&\frac{1}{4} \partial_{k}\partial_{l} \hat{n} (\br) \\
\hat{W}_l (\br) &=& \sum_{i=1}^N \delta (\br-\br_i) \Big( \sum_{j \neq i}^N \partial_{l} w(|\br-\br_j|) \Big)
\eea
represent the momentum-stress tensor and internal force density
of the system \cite{Tokatly}. Here $\overleftarrow{\partial}_{k,i}$ and $\overrightarrow{\partial}_{l,i}$ are partial derivatives
of particle $i$ that act to the left and right respectively.
If we replace $n([v],\br t)$ in eq.(\ref{InitialSturmLiouville}) by a {\em given} density $n$ subject to the 
following conditions on the initial density and its time-derivative at time $t_0$ 
\bea
n(\br t_0) &=& \braket{\Psi_0| \hat{n}(\br)}{\Psi_0} \label{initdens} \\
\partial_t n(\br t_0) &=&- \braket{\Psi_0| \nabla \cdot \hat{j}(\br)}{\Psi_0} \label{initdensderv}
\eea
where $\hat{j}(\br)$ is the usual current-density operator with 
\be
\hat{j}_k(\br) = \frac{1}{2 \imagi} \left( \sum_{i=1}^{N}  \delta (\br -\br_i) \overrightarrow{\partial}_{k,i} - 
 \overleftarrow{\partial}_{k,i} \delta (\br -\br_i)  \right),
\ee
then eq.(\ref{InitialSturmLiouville}) becomes 
\begin{eqnarray}
\label{InitialSturmLiouville2}
 -\nabla  \cdot  \left[ n(\br t) \nabla v(\br t) \right]   =   q([v], \br t) - \partial^{2}_t n(\br t) .
\end{eqnarray}
This is a nonlinear equation for $v$ which needs to be solved with
specified boundary conditions (this amounts to fixing a gauge for $v$).
If we propagate the TDSE with initial state $\ket{\Psi_0}$ and with a potential $v$ that is a solution to
eq.(\ref{InitialSturmLiouville2}) then for this potential clearly also the local force equation (\ref{InitialSturmLiouville}) will be satisfied with the same
initial conditions (\ref{initdens}) and (\ref{initdensderv}).
Subtracting 
eq.(\ref{InitialSturmLiouville2}) from (\ref{InitialSturmLiouville})
then yields the equation 
\be
\partial_t^2 \rho (\br t) - \nabla \cdot [ \rho (\br t) \nabla v (\br t) ] = 0
\label{rho}
\ee
for the density difference $\rho (\br t) = n([v],\br t)- n(\br t)$ with
initial conditions $\rho (\br t_0)=\partial_t \rho (\br t_0)=0$. Note, that the solution has further to fulfill $\int d \br \rho(\br t) = 0$.
The unique solution of eq.(\ref{rho}) with these initial conditions is $\rho (\br t) =0$ and hence $n(\br t)=n([v],\br t)$, i.e. the density in 
eq.(\ref{InitialSturmLiouville2}) is identical to the one that is obtained from time-propagation of the TDSE with the solution of eq.(\ref{InitialSturmLiouville2}). 
By making different choices for the density $n(\br t)$ in eq.(\ref{InitialSturmLiouville2}) we can deduce some important consequences of this result.
If we choose $n(\br t) = n([u],\br t)$ to be the density obtained by time-propagation of the TDSE with potential $u$ and the same initial state $\ket{\Psi_0}$
then we must have $n([u],\br t)=n([v],\br t)$ where $v$ is a solution of eq.(\ref{InitialSturmLiouville2}). The uniqueness of a 
potential $v$ for a given density $n$ (the Runge-Gross theorem) is thus equivalent to the uniqueness of the solution of eq.(\ref{InitialSturmLiouville2}), i.e. $u=v$.
If we choose $n(\br t)$ to be the density obtained by solving a TDSE for a system with different 
two-particle interactions $\hat{W}'$ and with different initial state $\ket{\Phi_0}$ then the existence of a solution to eq.(\ref{InitialSturmLiouville2}) 
implies that the same density can be reproduced by a potential $v$ in our system with interaction $\hat{W}$ and initial state $\ket{\Psi_0}$,
i.e. it is $v$-representable in our system. For the special case $\hat{W}=0$ this amounts to reproducing the density of an interacting system
within a non-interacting system, which is known as the Kohn-Sham construction and forms the basis of virtually all applications of TDDFT.
The key question, which is crucial for the whole foundation of TDDFT, is thus whether a solution to eq.(\ref{InitialSturmLiouville2}) 
is unique and exists. Existence and uniqueness have indeed been established \cite{RG,vanLeeuwen} under the restriction that the 
potential $v$ is Taylor-expandable around the initial time. The main goal of this paper is to present a proof that lifts this restriction.

\begin{figure}
\includegraphics[width=0.45\textwidth]{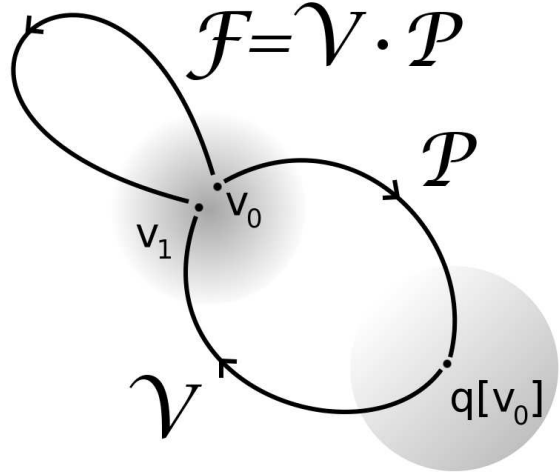} 
\caption{The potential-potential mapping $\mathcal{F}$ of eq.(\ref{mapF}) as composition of the mappings $\mathcal{P}$ and $\mathcal{V}$.}
\label{picture}
\end{figure}

Before going into details we first give the main idea.
For a potential $v_0 (\br t)$ and initial state $\ket{\Psi_0}$
at time $t_0$ we can propagate the
TDSE in a given time interval $[t_0,T]$ and construct the function $q([v_0], \br t)$ from eq.(\ref{qfunction}). This provides a mapping 
\begin{equation}
 \mathcal{P}: v_0 \mapsto q[v_0]
\end{equation}
(see fig.\ref{picture}). 
Let us fix a density $n(\br t)$ subject to the conditions (\ref{initdens}) and (\ref{initdensderv}) in terms of the initial state $\ket{\Psi_0}$. 
Then with the inhomogeneity $q[v_0]-\partial_t^2 n$
we can solve eq.(\ref{InitialSturmLiouville2}) for a new potential $v_1$, i.e.
\begin{eqnarray}
\label{InitialSturmLiouville3}
 -\nabla  \cdot  \left[ n(\br t) \nabla v_1(\br t) \right]   =   q([v_0], \br t) - \partial^{2}_t n(\br t) 
\end{eqnarray}
with given boundary conditions (this amounts to fixing a gauge for $v_1$). We point out that this is now a linear Sturm-Liouville problem parametrically 
dependent on time and as such not an evolution equation.
This provides us with a second map 
\begin{equation}
 \mathcal{V} : q[v_0] \mapsto v_1
\end{equation}
(see fig.\ref{picture}). The
combined map 
\be
\mathcal{F} [v_0] =(\mathcal{V} \circ \mathcal{P}) [v_0] = v_1
\label{mapF}
\ee
maps our original potential $v_0$ to a new one $v_1$. If for some potential $v$ we have
$v=\mathcal{F}[v]$, i.e. $v$ is a fixed point of the mapping $\mathcal{F}$, then we satisfy 
eq.(\ref{InitialSturmLiouville2}). Consequently, the question whether a solution to
eq.(\ref{InitialSturmLiouville2}) exists and is unique is equivalent to the question whether
the mapping $\mathcal{F}$ has a unique fixed point. This is exactly what we will show in this
work.
Our proof is based on the following inequality
\begin{eqnarray}
\label{step3}
\| \mathcal{F} [v_1] - \mathcal{F} [v_0] \|_{\alpha} \leq a \|v_1 - v_0\|_{\alpha}
\end{eqnarray}
with $a < 1$ and where $\| \cdot \|_{\alpha}$ is a norm dependent on a positive parameter $\alpha$ on the space of potentials. 
Comparable norms are commonly used in the solution of initial-value problems \cite{PartDiff}. This inequality can be derived in two steps. In the first step we prove that if two potentials are close in norm then also the observables $O ([v], t)$ calculated from
them are close. In particular for $\hat{O}=\hat{q}(\br)$ we prove
\be
\| q [v_1] - q [v_0] \|_{\alpha} \leq \frac{C}{\sqrt{\alpha}} \| v_1 - v_0 \|_{\alpha} 
\label{step1}
\ee
with $C$ a positive constant. In the second step we prove that 
\be
\| \mathcal{F}[v_1] - \mathcal{F}[v_0] \|_{\alpha} \leq D \| q [v_1] - q [v_0] \|_{\alpha}
\label{step2}
\ee
for a positive constant $D$.
The combination of these two statements then yields eq.(\ref{step3}) where 
$a=CD/\sqrt{\alpha}$ and we can choose $\sqrt{\alpha} > CD$.
It remains to prove Eqs.(\ref{step1}) and (\ref{step2}).

Before we do so, we point out conditions on the set of densities and potentials and some details of the mathematics involved.  All derivatives and solutions of the differential equations are defined in the weak sense, 
i.e. distributionally \cite{PartDiff}.
In order for the Sturm-Liouville boundary value problem of mapping $\mathcal{V}$ to be {mathematically} well defined we have to restrict our considerations on a  finite region $\Omega \subset \mathbf{R}^3$ with boundary $\partial \Omega$. 
This is not a serious restriction since this region can be arbitrarily large{, e.g. as big as the laboratory or the solar system,} and we can approach the continuum limit to any accuracy that
we desire.
Note that this restriction is also needed in the original {existence} proof \cite{vanLeeuwen} as was discussed in \cite{Math}. {The Runge-Gross theorem \cite{RG} also uses the Sturm-Liouville equation. As was already pointed out in \cite{Rajagopal}  and discussed to some extent in \cite{GrossKohn90} the proof needs further asymptotic conditions on the densities and potentials. These asymptotic requirements are most easily met by choosing a finite volume.} 
The domain of the mapping $\mathcal{F}$ is the Sobolev space of distributionally one-times spatially differentiable 
square-integrable potentials \cite{Math, PartDiff}. This space is similar to the well-known domain of, e.g. the Laplace operator {$\nabla^2$}.  As the domain is dense in the set the square-integrable functions we can uniquely extend
our mapping via Cauchy-sequences to this set. In this way we can then also converge to Coulombic external potentials since they are square integrable, similarly to the case of the Laplace operator.  The space of the potentials \cite{mage} is isomorphic to the Banach space of
strongly measurable functions from $[t_0,T]$ onto the Hilbert space of square-integrable functions in space 
with the norm $\mathrm{ess\; sup}_{t_0 \leq t \leq T} \| v(t) \|^2 < \infty$, where $\| v(t) \|^2 = \int_{\Omega} d\br \, v(\br t)^2$ and $\mathrm{ess \; supp}$ denotes 
the essential supremum \cite{PartDiff}. 
If the density $n(\br t)$ in eq.(\ref{InitialSturmLiouville3}) is not divergent at any point then for all possible $v$ under considerations $n(\br t)\nabla v(\br t)$ is square-integrable. 
If we then require the second 
time-derivative of the density $\partial_t^2 n(\br t) ${ to be in the dual to the Sobolev space of potentials \footnote{A thorough discussion of these mathematical details can be found in \cite{mage}}, i.e. a requirement which is slightly weaker than square-integrability,} 
then from eq.(\ref{InitialSturmLiouville}) we see that the divergence of the local forces 
are also {in this dual space}.
We thus restrict ourselves to finite densities {having a second time-derivative in this dual space. Similar restrictions have to be posed in \cite{vanLeeuwen} as has been discussed in \cite{vanLeeuwenIJMP, Math}. With these restrictions the Sturm-Liouville problem  (\ref{InitialSturmLiouville3}) can be solved uniquely. The stronger restriction of square-integrable second time-derivatives of the densities and square-integrable internal force divergences is usually assumed to hold for all physical wave-functions \cite{Maitraetal}.}

We start by deriving inequality eq.(\ref{step1})
for the mapping $\mathcal{P}$ of Fig.(\ref{picture}). The main ingredient is the fundamental theorem of
calculus \cite{Griffel}. Suppressing the time-arguments we can write
\be
\label{FunDerivative}
O [v_1]-O [v_0] = \int_0^1 d\lambda \, \frac{dO}{d\xi} [v_0 + \xi (v_1-v_0)] |_{\xi=\lambda} .
\ee
for any operator expectation value $O([v],t)$.
In the case that we take $\hat{O}=\hat{q}(\br)$ this equation yields
\bea
\lefteqn{q([v_1],\br t)-q([v_0],\br t) } \nonumber \\
&=& \int_{t_0}^t dt' \int_{\Omega} d\br' \chi (\br t, \br' t') (v_1 (\br' t') - v_0 (\br' t'))
\label{linresp}
\eea
where we defined
\be
\chi (\br t, \br' t') = -i \,\int_0^1 d\lambda 
\braket{\Psi_0| [\hat{q}_{H_{\lambda}} (\br t), \hat{n}_{H_{\lambda}}(\br' t')]}{\Psi_0} 
\label{linresp2}
\ee
and $\hat{O}_{H_{\lambda}}$ is the operator $\hat{O}$ in the Heisenberg representation with respect to Hamiltonian
$H_{\lambda}$ with potential $v_{\lambda}=v_0 + \lambda (v_1-v_0)$. Equation (\ref{linresp2}) can be obtained
directly from the TDSE by evaluating the expectation value $q([v_\xi], \br t)$ to first order in $\xi$
around $\lambda$ which amounts to linear response theory \cite{TDDFT}. {In the following we fix $v_1$ and $v_0$ and derive an approximation for the induced difference in the local-force divergences. The linear response kernel $\chi$ depends on this choice of potentials, i.e. $\chi = \chi[v_0,v_1]$.}
{Equation} (\ref{linresp}) {has the form}
\be
f(\br t) = \int_{t_0}^t dt' \int_{\Omega} d\br'  \chi (\br t, \br' t') g(\br' t') = (\chi g) (\br t) .
\ee
One can then derive that
\be
\| f (t) \|^2 \leq C(t)^2 \int_{t_0}^t dt' \, \| g(t') \|^2. 
\label{ineq1}
\ee
In this expression the function $C(t)$ (the operator norm) is defined as
\be
C (t)^2 = \sup_{g \neq 0} \frac{\| (\chi g) (t) \|^2}{\int_{t_0}^t dt'  \| g(t') \|^2 }.
\label{opnorm}
\ee
{Note, that we have fixed $\chi = \chi[v_0,v_1]$ while ranging over different functions $g$. We assume $C(t)$ to be finite, i.e. the operator norm of $\chi$ to exist. Due to the dependence of $\chi$ on the potentials chosen also $C(t)$ depends on this choice.} Using a Cauchy-Schwarz inequality the function $C(t)$ can be shown to satisfy
\be
C(t)^2 \leq \max_{t' \in [t_0,t]} \int_{\Omega} d\br d\br' Ê\chi (\br t, \br' t') ^2.
\ee
The integral on the right hand side of inequality (\ref{ineq1}) can be manipulated as follows
\bea
\int_{t_0}^t dt' \, \| g(t') \|^2 &=& \int_{t_0}^t dt' \, e^{-\alpha (t'-t_0)}  e^{\alpha (t'-t_0)}\| g(t') \|^2 \\
&\leq& \| g \|_{\alpha,t}^2 \int_{t_0}^t dt' e^{\alpha (t'-t_0)} \leq \| g \|_{\alpha,t}^2 \frac{e^{\alpha (t-t_0)}}{\alpha} \nonumber 
\eea
where $\alpha$ is an arbitrary positive number and where we defined
\be
\label{alphanorm}
\| g \|_{\alpha,t}^2 = {\mathrm{ess} \!\!\! \sup_{t' \in [t_0,t]}} \Big( \| g(t') \|^2 \, e^{-\alpha (t'-t_0)} \Big) .
\ee
With this result inequality (\ref{ineq1}) can be written as
\be
\| f \|_{\alpha,t}^2 \leq \frac{C(t)^2}{\alpha}  \| g \|_{\alpha,t}^2 .
\ee
We will consider functions on an arbitrarily large but finite interval $[t_0,T]$ with $T > t_0$. If we define
$\|Êf \|_{\alpha} = \| f \|_{\alpha,T}$ and $C=C(T)$ and apply this to eq.(\ref{linresp}) we obtain eq.(\ref{step1}).
This concludes the first part of the proof. The discussion so far was related to the mapping $\mathcal{P}$. Let us
now discuss the second mapping $\mathcal{V}$.\\
We consider the Sturm-Liouville operator $Q\! =\! -\nabla \!\cdot\! [n \nabla]$. By partial integration we find for two external potentials $v_0$ and $v_1$
within the standard inner product
\begin{eqnarray}
\label{selfadjoint}
\lefteqn{ \braket{v_0}{Qv_1} - \braket{Qv_0}{v_1} =} 
\\
 && \int_{\partial \Omega} \diff S \cdot n(\br t)\left(v_1(\br t) \nabla v_0(\br t) - v_0(\br t) \nabla v_1(\br t)   \right).  \nonumber
\end{eqnarray}
Hence, in order for $Q$ to be self-adjoint the boundary term has to vanish. If $n(\br t) \geq  \epsilon >0 $ in $\Omega$ we know that we find solutions to the general inhomogeneous Sturm-Liouville problem \cite{Math} with the boundary conditions $v=0$ on $\partial \Omega$, making $Q$ a self-adjoint operator. We then have an orthonormal set of eigenfunctions \cite{Griffel}. {If the density is zero at the boundary the same conclusion holds under the assumption of $n^{-1}$ being square-integrable \cite{mage}. In general a boundedness condition at the edge is assumed to single} out a unique set
(a famous example is the Legendre equation which would correspond to an operator $Q$ with density $n(x)=(1-x^2)$ on $[-1,1]$ ).
For an extensive discussion of these issues in the one-dimensional case see \cite{Bailey}. 
In general we have a set of orthonormal eigenfunctions $\{ \phi_{i}(\br t) \}$ with $Q \phi_i = \lambda_i \phi_i $ and positive eigenvalues $0 \leq \lambda_0 < \lambda_1 < ...$, since $
 \lambda_i = \braket{\phi_i}{Q \phi_i} = \braket{\nabla \phi_i}{n \nabla \phi_i} \geq 0$. The eigenfunction to the eigenvalue zero is $\phi_0 =c(t)$.
If we now subtract eq.(\ref{InitialSturmLiouville3}) for $v_1 =\mathcal{F}[v_0]$ from the one for $v_2 =\mathcal{F}[v_1]$ we obtain
\be
\label{InitialSturmLiouville4}
 -\nabla  \cdot  \left[ n(\br t) \nabla (v_2(\br t) -v_1 (\br t) ) \right]   =   q([v_1], \br t) -  q([v_0], \br t) .
\ee 
We can then expand $v_2-v_1$  and $\zeta=q[v_1]-q[v_0]$ in terms of the eigenfunctions of $Q$, i.e, $(v_2-v_1) (\br t) = \sum_{i=0}^{\infty} u_i(t) \phi_i(\br t)$ and $\zeta(\br t) = \sum_{i=0}^{\infty} \zeta_i(t) \phi_i(\br t)$.
Since $\zeta$ is a divergence $\int \diff^3 r \, \zeta (\br t) = 0$ and therefore the constant function $\phi_0$ does not contribute to the expansion. Likewise the gauge fixing allows us to exclude $\phi_0$ from the expansion of $v_2-v_1$. Inserting the expansions into eq.(\ref{InitialSturmLiouville4}) yields 
\begin{equation}
 u_{i}(t) = \frac{\zeta_i(t)}{\lambda_i(t)}
\end{equation}
where $\lambda_i(t) > 0$ and with this we have
\begin{eqnarray}
\label{EigenfctEstimation}
\| v_2 (t) - v_1 (t) \|^2 &=&  \sum_{i=1}^{\infty} \left| \frac{\zeta_i(t)}{\lambda_i(t)} \right|^2  \leq \frac{1}{\lambda_1(t)^2} \sum_{i=1}^{\infty} \left| \zeta_i(t) \right|^2 
\nonumber \\
&=& \frac{1}{\lambda_1(t)^2} \| q[v_1](t) - q[v_0](t) \|^2. 
\end{eqnarray}
If we now multiply eq.(\ref{EigenfctEstimation}) with $e^{-\alpha (t-t_0)}$ and take a maximum over the interval $[t_0,T]$ we arrive at
eq.(\ref{step2}) in which $D^2 = \max_{t \in I}\{ \lambda_1(t)^{-2}\}$. This then together with eq.(\ref{step1}) establishes our main inequality (\ref{step3}).

This equation can now be used to prove the uniqueness of a solution to eq.(\ref{InitialSturmLiouville2}). Suppose we would have two fixed point solutions $u$ and $v$, i.e.
$u=\mathcal{F}[u]$ and $v=\mathcal{F}[v]$. Then by choosing $\sqrt{\alpha}=2 \,CD$ {for this pair of potentials} in eq.(\ref{step3}) we find
\bea
\| v- u \|_{\alpha} = \| \mathcal{F}[v] - \mathcal{F}[u] \|_{\alpha} \leq \frac{1}{2} \| v - u \|_{\alpha}
\eea
from which we conclude {that $\| v- u \|_{\alpha}  = 0$ and thus we have} $u=v$. Hence, if a solution to  eq.(\ref{InitialSturmLiouville2}) exists then it is unique.
This conclusion is equivalent to the Runge-Gross theorem \cite{RG}. It states that a density $n([v],\br t)$ cannot be produced by
another potential $u$ starting from the same initial state. This is now proven without assumptions on the Taylor-expandability in time of the
potential. Suppose that the density of an interacting system is representable in a non-interacting system then this theorem guarantees
that the effective potential producing the same density in this system is unique. This establishes the uniqueness of a Kohn-Sham scheme.
\\
Let us now address the existence of a solution to eq.(\ref{InitialSturmLiouville2}). This is a $v$-representability question for a given density. 
We see from eq.(\ref{opnorm}) that the constant
$C=C(T)$ in eq.(\ref{step1}) is dependent on the response function $\chi$ and hence via eq.(\ref{linresp}) on potentials $v_0$ and $v_1$, i.e. $C=C[v_0,v_1]$. { We assume that a constant $C_{\textrm{sup}}=\sup_{v_0} C [v_0,\mathcal{F}[v_0]]$} exists when we range over all potentials $v_0$. 
{ÊWe now follow exactly the same reasoning as in the proof of the Banach fixed point theorem \cite{Griffel}.
Let $v_k=\mathcal{F}^k [v_0]$ denote the $k$-fold application of the mapping $\mathcal{F}$ on a given initial potential $v_0$
and choose $\sqrt{\alpha} > \, C_{\textrm{sup}}D$. Then eq.(\ref{step3}) with $a=C_{\textrm{sup}} D/\sqrt{\alpha}$ implies $\| v_{k+1} - v_{k} \|_{\alpha} \leq a^k \| v_1 - v_0 \|_{\alpha}$ which means that the $v_k$ are a Cauchy series.
Since the set of potentials is a Banach space with the norm (\ref{alphanorm}) \cite{PartDiff, mage} and therefore complete this series
converges to a unique $v$ , i.e. $v_k \rightarrow v$ for $k \rightarrow \infty$. 
According to our assumption the response function of eq.(\ref{linresp2}) exists and hence $q$ is functionally differentiable and 
consequently continuous as a functional of $v$.
Therefore $\lim_{k \rightarrow \infty} q[v_k] = q[v]$ which means that $v$ solves eq.(\ref{InitialSturmLiouville2})
and hence is a fixed point. This establishes the existence of a Kohn-Sham system corresponding to the density $n$ in eq.(\ref{InitialSturmLiouville2}) provided there is a supremum $\sup_{v_0} C [v_0,\mathcal{F}[v_0] ]$
when we range over potentials $v_0$ in a non-interacting system.
}

We have generalized the {uniqueness and existence} theorems of TDDFT to cover the case of external potentials that are not Taylor-expandable in time using a new formulation of the problem 
as a fixed point problem in terms of a suitably chosen norm. {
We showed that a potential generating a given density is unique provided that
the response function of eq.(\ref{linresp2}) has a finite operator norm and the second time-derivative of the density to obeys a slightly weaker condition than square-integrability. 
We further showed the existence of an external potential producing a given density provided there exists
an upper bound for the operator norm of all response functions between successive potentials of the iterative procedure.}
The main idea has several applications in other research fields. For instance, we will extend the ideas presented here to other density functional theories, e.g., current-density functional theory \cite{Vignale} and also study the convergence properties of optimized effective potential approaches \cite{Ullrich, vanLeeuwen3, vonBarth}. As the presented iteration scheme poses a way to deduce an external potential generating a given density, this could also prove useful for quantum optimal control theory \cite{OCT}.

\acknowledgments
Financial support by the Erwin Schr\"odinger Fellowship J 3016-N16 of the FWF (Austrian Science Fonds) and fruitful discussions with M.\ Penz are gratefully acknowledged.


\begin{thebibliography}{0}

\bibitem{TDDFT} 
  \Name{Marques M.A.L. , \etal }
  \Book{Time-Dependent Density Functional Theory}
  \Publ{Springer, Heidelberg}
  \Year{2006}.
  
  \bibitem{vanLeeuwenIJMP} 
\Name{van Leeuwen R.}
\REVIEW{Int. J. Mod. Phys. B}{15}{2001}{1969}. 

\bibitem{Marques} 
\Name{Marques M. A. L. and Gross E. K. U.} 
\Book{in The Electron Liquid Paradigm in Condensed Matter Physics, Proceedings of the International School of Physics ÒEnrico FermiÓ, Course CLVII} 
\Editor{GIULIANI G. F. and VIGNALE G.} 
\Publ{IOS Press, Amsterdam} 
\Year{2004}
\Page{127-167}.

\bibitem{RG} 
\Name{Runge E. and Gross E.K.U}
\REVIEW{Phys. Rev. Lett.}{52}{1984}{997}. 

\bibitem{vanLeeuwen}
  \Name{van Leeuwen R.}
  \REVIEW{Phys. Rev. Lett.}{82}{1999}{3863}. 

\bibitem{Maitraetal} 
\Name{Maitra N.T., Todorov T.N., Woodward C. and Burke K.}
\REVIEW{ Phys. Rev. A}{81}{2010}{042525}.

\bibitem{RuggenthalerPRA} 
\Name{Ruggenthaler M., Penz M. and Bauer D.}
\REVIEW{Phys. Rev. A}{81}{2010}{062108}.

\bibitem{TokatlyArXiv} 
\Name{Tokatly I.V.}
 \REVIEW{Phys. Rev. B}{83}{2011}{035127}.
 
\bibitem{Tokatly} 
\Name{Tokatly I.V.}
\REVIEW{Phys. Rev. B}{71}{2005}{165104}.

\bibitem{PartDiff} 
\Name{Evans L.C.}
\Book{Partial Differential Equations}
\Publ{American Mathematical Society, Rhode Island} 
\Year{2010}.

\bibitem{Math} 
 \Name{Ruggenthaler M., Penz M. and Bauer D.}
\REVIEW{J. Phys. A: Math. Theor.}{42}{2009}{425207}.

{
\bibitem{Rajagopal}
\Name{Xu B.X. and Rajagopal A.K.}
\REVIEW{Phys. Rev. A}{31}{1985}{2682}.
}

{
\bibitem{GrossKohn90}
\Name{Gross E.K.U. and Kohn W.}
\REVIEW{Adv. Quant. Chem.}{21}{1990}{255 }.
}

{
\bibitem{mage} 
 \Name{Penz M. and Ruggenthaler M.}
\REVIEW{arXiv:1103.1983 preprint}{}{2011}{}.
}


\bibitem{Griffel} 
\Name{Griffel D.H.}
\Book{Applied Functional Analysis} 
\Publ{Ellis Horwood Ltd, Chichester} 
\Year{1985}.

\bibitem{Bailey}  
\Name{Bailey P.B., Everitt W.N., and Zettl A.}
\REVIEW{ACM Trans. Math. Software}{21}{2001}{143} . 

\bibitem{Vignale} 
\Name{Vignale G.}
\REVIEW{Phys. Rev. B}{70}{2004}{201102 (R)}.

\bibitem{Ullrich} 
\Name{Wijewardane H.O. and Ullrich C.A.} 
\REVIEW{Phys. Rev. Lett.}{100}{2008}{056404}. 

\bibitem{vanLeeuwen3}
\Name{van Leeuwen R.}
\REVIEW{Phys. Rev. Lett.}{76}{1996}{3610}.

\bibitem{vonBarth}
\Name{von Barth U., Dahlen N.E., van Leeuwen R. and Stefanucci G.}
\REVIEW{Phys. Rev. B}{72}{2005}{235109}. 

\bibitem{OCT}  
\Name{Peirce A. P. , Dahleh M.A. and Rabitz H.}
\REVIEW{Phys. Rev. A}{37}{1988}{4950}.

\end{thebibliography}
\end{document}